\documentclass[nofootinbib,amssymb]{revtex4}
\usepackage{graphicx}
\def\dd{\displaystyle}
\begin{document}
\title{\bf A Renormalized perturbation theory for problems with nontrivial boundary conditions or backgrounds in two space-time dimensions}
\author{Reza Moazzemi}
\author{Abdollah Mohammadi}
\thanks{Present address: Department of Physics, Shiraz University, Shiraz 71345, Iran}\email{amohammadi@shirazu.ac.ir}
\author{Siamak S. Gousheh}%

\affiliation{%
Department of Physics, Shahid Beheshti University  G. C., Evin, Tehran
19839, Iran
}%
\date{\today}

\begin{abstract}
In this paper we discuss the effects of nontrivial boundary
conditions or backgrounds, including non-perturbative ones, on the renormalization program for
systems in two dimensions. Here we present an alternative renormalization procedure such that these non-perturbative conditions can
be taken into account in a self-contained and, we believe, self-consistent
manner. These conditions have profound effects on the properties
of the system, in particular all of its $n$-point functions. To be
concrete, we investigate these effects in the $\lambda \phi^4$ model in two dimensions and show that the mass counterterms turn out to be proportional to the Green's  functions which have nontrivial position dependence in these cases. We then compute the difference between the mass counterterms
in the presence and absence of these conditions. We find that in
the case of nontrivial boundary conditions this difference is minimum between the boundaries and
infinite on them. The minimum approaches zero when the boundaries go to infinity. In the case
of nontrivial backgrounds,
we consider the kink background and show that
the difference is again small and localized around the kink.
\end{abstract}

\maketitle

\section{Introduction}

The procedure of the renormalization with no nontrivial
backgrounds or boundary conditions, is standard and has been available for over half a century \cite{bogoliubov}. However there has been much less investigations done on the renormalization programs for the system which are subject to nontrivial boundary conditions or include nontrivial backgrounds \cite{Jackiw1}, including non-perturbative ones such as solitary waves or solitons \cite{Coleman0}. It is worth mentioning that for the purposes of this paper the distinction between the solitary waves and solitons is unimportant we use them interchangeably. The usual renormalization procedure for systems with such non-perturbative conditions has been either identical to the analogous free cases or has included slight modifications (see for example \cite{dashen,raja,reb97,Fosco,Baron1}). An important example for the case of nontrivial boundary conditions is the calculation of radiative corrections to the Casimir effect. In this case some authors have used free renormalization programs in which the counterterms are directly imported from the free cases or at most supplemented with delta functions on the boundaries \cite{Baron2}. An important example for the case of nontrivial backgrounds is the calculation of quantum corrections to the mass of solitons. In this case the predominant practice has been to use the free renormalization procedure.
The main issue that we want to discuss in this paper is that the presence
of these non-perturbative conditions has profound effects on all physical properties of the system, and the ensuing renormalization procedure. That is, these conditions are an integral part of the overall structure and
properties of the theory and obviously cannot be ignored or even
taken into account perturbatively. In
fact we believe that the solution to the problem could be
self-contained and the renormalization procedure be done
self-consistently with the nature of the problem.

An additional justification supporting our proposed method is the
fact that the presence of these non-perturbative conditions break the
translational symmetry of the system. For example in the presence
of soliton this occurs when we fix the position of the solitons.
Obviously the breaking of the translational symmetry has many
manifestations. Most importantly, all the \emph{n}-point functions
of the theory will have in general nontrivial position dependence
in the coordinate representation. The procedure to deduce the
counterterms from the \emph{n}-point functions in a renormalized
perturbation theory is standard. This, as we shall show, will
lead to uniquely defined position dependent
counterterms which need to be fixed only at one spatial point. Therefore, the radiative corrections to all the
input parameters of the theory, will be in general position
dependent. Therefore, we believe the information about the
nontrivial boundary conditions or position dependent backgrounds is
carried by the full set of \emph{n}-point functions, the resulting
counterterms, and the renormalized parameters of the theory.

In this paper we set up alternative renormalized perturbation theories for
$\phi^4$ model in two dimensions for two different cases: First,
the $\phi^4$ theory with Dirichlet boundary conditions. One of the
most important application of such theory is the calculation of
the radiative corrections of the Casimir effect. Second, $\phi^4$
theory in the spontaneously broken symmetry phase with the static
solitary wave (the kink) as the background. The most common use this theory is the
calculation of the quantum corrections to the mass of the kink. We
calculate the mass counterterms in these cases. In the both cases
we will inevitably obtain position dependent counterterms. For the
first one, our results show that the main difference between the counterterms in the presence and absence of boundary conditions is for positions which are about a
Compton wavelength away from the walls, although it has a small value at other places. When the boundaries go to infinity, our counterterm approaches the free one supplemented by delta functions at the boundaries. This is precisely the modified counterterm chosen by \cite{Fosco} for all values of distance between the plates. In this sense, we believe, their counterterm is only an approximation to ours. In the second case, we explicitly show
that the difference between our mass counterterm in the presence of the kink and the free one
is small and localized around the central position of the kink.

\section{The counterterms in the presence of boundary conditions}

In this section we present our alternative approach to the
renormalization of scalar field confined between two points in two
space-time dimensions. The Lagrangian density for a real scalar
field with $\phi^4$ self-interaction is:
\begin{equation}\label{e1}
  {\cal L}(x)
  =\frac{1}{2}[\partial_{\mu}\varphi(x)]^{2}-\frac{1}{2}m_0^{2}\varphi(x)^{2}
  -\frac{\lambda_{0}}{4!}\phi(x)^{4}
\end{equation}
where $m_{0}$ and $\lambda_{0}$ are the bare mass and bare
coupling constant, respectively. The Lagrangian after rescaling
the field $\phi=Z^{1/2}\phi_{r}$, where $Z$ is called the
field strength renormalization, and the standard procedure for
setting up the renormalized perturbation theory, becomes (see for
example \cite{Peskin}),
\begin{eqnarray}\label{e17}
   &&{\cal L}(x)=\frac{1}{2}[\partial_{\mu}\phi_{r}(x)]^{2}-\frac{1}{2}m^{2}
   \phi_{r}(x)^{2}-\frac{\lambda}{4!}\phi_{r}(x)^{4}\qquad\nonumber\\&&\hspace{2cm}+
   \frac{1}{2}\delta Z[\partial_{\mu}\phi_{r}(x)]^{2}
   -\frac{1}{2}\delta m^2\phi_{r}(x)^{2}-\frac{\delta\lambda}{4!}\phi_{r}(x)^{4},
\end{eqnarray}
where $\delta m^2,\delta\lambda,\delta Z$ are the counterterms,
and $m$ and $\lambda$ are the physical mass and physical coupling
constant, respectively. We should mention that
in these 1+1 dimensional problems, one usually chooses a minimal
renormalization scheme defined at all loops by
\cite{dashen,raja,reb97}
\begin{equation}
\delta Z  = 0\quad,\quad \delta\lambda=0 \quad\mbox{ and}\quad m_0
^2  = m^2  - \delta m^{2}.
\end{equation}
The sufficiency of these conditions is
supported by the fact that for any theory of a scalar field in two
dimensions with nonderivative interactions, all divergences that
occur in any order of perturbation theory can be removed by
normal-ordering the Hamiltonian \cite{coleman}. Complete calculations of all the counterterms in higher dimensions are obviously more complicated and will not be attempted here.

In this problem we are to impose appropriate boundary conditions
on the field at the end points. Obviously the presence of
nontrivial boundary conditions breaks the translational invariance
and hence momenta will no longer be good quantum numbers.
Therefore we find it easier to impose the renormalization
conditions in the configuration space. For example, the standard
expression for the two-point function in the renormalized
perturbation theory is,
\begin{eqnarray}\label{2pfun.}
    \langle\Omega
    |T\{\phi(x_1)\phi(x_2)\}|\Omega\rangle=\lim_{T\to\infty(1-i\epsilon)}
    \frac{\langle0|\int{\cal D}\phi\phi(x_1)\phi(x_2)e^{i\int_{-T}^T
    {\cal L}(x) d^4x }|0\rangle}{\langle0|\int{\cal D}\phi e^{i\int_{-T}^T {\cal
    L}(x)
    d^4x}|0\rangle}.
\end{eqnarray}
Since the birth of quantum field theory, as far as we know, the
assertion has always been that the above expressions can be expanded
systematically when the problem is amenable to perturbation theory.
For example, in the context of renormalized perturbation theory, as
indicated in Eq.(\ref{e17}), we can symbolically represent the first
few terms of the perturbation expansion of Eq.(\ref{2pfun.}) by
\begin{equation}\label{e18:renor.con1.}
   \raisebox{-4.mm}{\includegraphics[width=1.5cm]{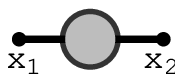}}=\raisebox{-2.5mm}{\includegraphics[width=1.5cm]{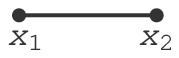}}
   +\raisebox{-2.7mm}{\includegraphics[width=1.1cm]{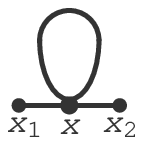}}+\raisebox{-3.5mm}{\includegraphics[width=1.5cm]{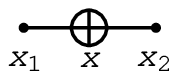}}+\dots.
\end{equation}
where \raisebox{-2.5mm}{\includegraphics[width=1.3cm]{16}} refers
to the appropriate counterterm. It is obvious that the above
expression represents a systematic perturbation expansion, and
most importantly, all of the propagators on the right hand side
should be the ones appropriate to the problem under consideration.
That is, they should have the same overall functional form as the
first term. An integral over space-time is implicitly assumed in
the above expression, and the final result is obviously
$x$-independent. Our first renormalization condition is equivalent
to the usual one which states that the exact propagator should
equal the the propagator represented by the first term in
(\ref{e18:renor.con1.}) close to its pole. This implies the second and third
diagrams should cancel each other out in the lowest order, and
this in turn implies the cancelation of the UV divergences to that
order. The presence of an space-time integral clouds the central issue at
this point. However considering the higher order diagrams, one can
easily conclude that the cancelation of the two aforementioned
terms should occur locally. This implies that the counterterms
will in general turn out to be position dependent. Therefore we
have
\begin{equation}\label{e20:counterterms}
   \delta m^2(x)=\frac{-i}{2}\raisebox{-2.4mm}{\includegraphics{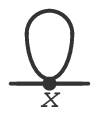}}
   =\frac{-\lambda}{2}G(x,x).
\end{equation}
Here $G(x,x')$ is the propagator of the real scalar field and
$x=(t,z)$. Obviously the counterterms automatically
incorporate the boundary conditions and are position dependent,
due to the dependence of the two and four-point functions on such
quantities.  An interesting point is that if the propagator for
the loop part of Eq. (\ref{e18:renor.con1.}) is taken to be the
free propagator, i.e. the one with no non-trivial boundary
conditions, then the counterterm will turn out to be exactly the
free one. However, we have reservations about this procedure. That is, we believe a self-contained and self-consistent procedure is one in which
all of the propagator segments of the above expansion are of the
same form, i.e. the one represented by the first diagram. This
is strengthened by the symmetry breaking argument presented in the
introduction. Consequently the exact functional dependence of
$\delta m^2(x)$ can be completely determined by the theory. That is,
the overall structure of the renormalization conditions
such as above, and the counterterms appearing in them could be
determined solely from within the theory, and there is no need for example to
import them from the free case. The significance of the
difference between the counterterms can be illuminated by the following relationship
\begin{equation}
m_0^2  = m_{\rm{free}}^2  - \delta m_{\rm{free}}^{2}=m_{\rm{bound}}^2  - \delta m_{\rm{bound}}^{2}.
\end{equation}
It is extremely important to mention that it is sufficient to fix $m_{\rm{bound}}$ only at one value of $x$.

\begin{figure}[th]\begin{center} \centerline{\includegraphics[width=11cm]{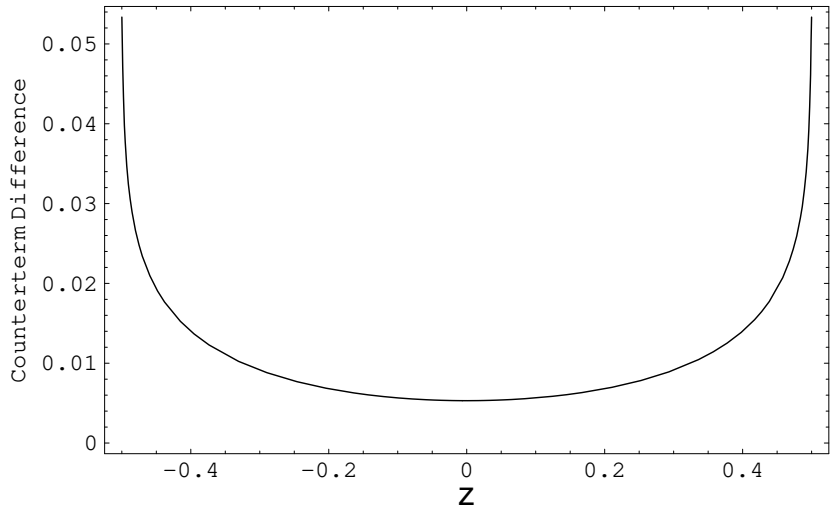}}
\caption{\small The difference between the counterterms for the $\lambda \phi ^4$ Dirichlet problem: $ \delta m_{\rm{bound}}^2(x)-\delta m_{\rm{free}}^2$, for mass $m=1$;
$a=1(\mbox{m}^{-1})$ and $\lambda=0.1(\mbox{m}^2)$. Please note that this difference is significant only close to the boundaries. At the boundaries this difference diverges logarithmically. When the plate separations goes to infinity this difference  approaches zero for all finite $z$. }\label{fig1}
\end{center}\end{figure}

Now we study very briefly the specific effects of confinement of
the system to a finite size by imposing Dirichlet boundary
condition. In order to do this we have to compare the difference
between the counterterms of free space and bounded one in the interval
$\left[-\frac{a}{2},\frac{a}{2}\right]$:
 \begin{eqnarray}\label{coun.diff}
   \delta m_{\rm{bound}}^2(x)-\delta m_{\rm{free}}^2=\frac{-\lambda}{2}\left[G_{\rm{bound}}(x,x)-G_{\rm{free}}(x,x)\right]
\end{eqnarray}
For the free space we have $G_{\rm{free}}(x,x')=\int
\frac{d^2k}{(2\pi)^2} \frac{e^{-ik(x-x')}}{k^2-m^2+i\epsilon }$
which in the Euclidean space leads to
\begin{equation}\label{freecount}
   G_{\rm{free}}(x,x)=\int \frac{d\omega}{2\pi} \int \frac{dk}{2a} \frac{1}{\omega'^2+\frac{k^2\pi^2}{a^2}}=\frac{1}{4\pi }\int \frac{d\omega}{\omega'}.
\end{equation}
where $\omega'^2=\omega^2+m^2$. For the bounded problem we have
the following expression for the Green's function in the two
dimensional Euclidean space
\begin{eqnarray}\label{e24:G.Func.}
    G_{\rm{bound}}(x,x')=\frac{2}a\int\frac{d\omega}{2\pi}e^{\omega(t'-t)}\sum_{n}
   \frac{\sin\left[k_{n}(z+\frac{a}{2})\right]\sin
   \left[k_{n}(z'+\frac{a}{2})\right]}{\omega'^{2}+k_{n}^{2}}.\nonumber\\
\end{eqnarray}
where $k_{n}=\frac{n\pi}{a}$ is the momentum perpendicular to
plates.
 Setting $x'=x$ and using some trigonometric identities, we get
 \begin{eqnarray}\label{coun.diff2}
   \delta m_{\rm{bound}}^2(x)-\delta m_{\rm{free}}^2=\frac{-\lambda}{8a\pi}\int d\omega\left\{\frac{a\omega'\coth(a\omega')-1}{\omega'^2}
   -2\sum_{n=1}^\infty \frac{\cos\left[2k_{n}(z+\frac{a}{2})\right]}{\omega'^{2}+k_{n}^{2}}-\frac{a}{\omega'} \right\}.
\end{eqnarray}
The summation term in the above equation can be written in
terms of hypergeometric functions as follows
\begin{eqnarray}\label{coun.diff2...}
 &&\sum_{n=1}^\infty \frac{\cos\left[2k_{n}(z+\frac{a}{2})\right]}{\omega'^{2}+k_{n}^{2}}=\frac{iae^{-\frac{2i\pi z}{a}}}{4\omega'(\pi^2+a^2\omega'^2)}\nonumber\\
 &&\Bigg\{(ia\omega'-\pi)\left[ {_2F_1}\left(1,1+\frac{ia\omega'}{\pi},2+\frac{ia\omega'}{\pi},-e^{-\frac{2i\pi z}{a}}\right)+{e^{\frac{4i\pi z}{a}}} _2F_1\left(1,1+\frac{ia\omega'}{\pi},2+\frac{ia\omega'}{\pi},-e^{\frac{2i\pi z}{a}}\right)\right]\nonumber\\&&+(ia\omega'+\pi) \left[ _2F_1\left(1,1-\frac{ia\omega'}{\pi},2-\frac{ia\omega'}{\pi},-e^{-\frac{2i\pi z}{a}}\right)+{e^{\frac{4i\pi z}{a}}} _2F_1\left(1,1-\frac{ia\omega'}{\pi},2-\frac{ia\omega'}{\pi},-e^{\frac{2i\pi z}{a}}\right)\right]\Bigg\}.
\end{eqnarray}
It is important to note that the counterterm for the bounded case
has only a logarithmic ultraviolet divergence, while the free one
has an additional infrared divergence in the massless case. The
ultraviolet divergences exactly cancel each other for all cases,
while the infrared divergence in the massless case remains. The
resulting difference between the counterterms, Eq.
(\ref{coun.diff2}), is illustrated in Fig. \ref{fig1} for $m=1$;
$a=1(\mbox{m}^{-1})$ and $\lambda=0.1(\mbox{m}^2)$, where m denotes the mass of the scalar particle.
This figure shows that the difference between the counterterms in
the free and bounded cases is minimum in the middle of the plates,
and infinite on the plates. It is interesting to note that, for
$m\neq0$, as $a\to\infty$, this difference approaches zero for all
finite $z$. We can  compare our results with Refs. \cite{Fosco} in
which the authors attempt to include finite size effects in their
renormalization procedure. The counterterm which they use is just
the free counterterm between the plates and delta functions on the
plates for all plate separations. This is identical to what we have only in the limit of infinite plate separation. We believe the counterterm that we have obtained within
our theory is the one which is appropriate to this problem and
their counterterm is only an approximation to ours. We have
obtained the the first order radiative correction to the Casimir
energy using our counterterm \cite{Gousheh1,Gousheh2}.

\section{The counterterms in the presence of nontrivial backgrounds}

In this section we study the counterterms appropriate for problems
with nontrivial backgrounds. One interesting background is the
$\phi^4$ kink. This background has nontrivial boundary values as
well as nontrivial spatial variations. We start with the
Lagrangian density for a neutral massive scalar field, within
$\phi^4$ theory, appropriate for the spontaneously broken symmetry
phase in 1+1 dimensions:
\begin{equation}\label{1}
{\cal L} =\frac{1}{2}\left(\frac{{\partial \phi }}{{\partial
t}}\right)^2 - \frac{1}{2}\left(\frac{{\partial \phi }}{{\partial
z}}\right)^2 -U[\phi(x)],
\end{equation}
where $U[\phi]=\frac{\lambda'_ 0}{4}\left( \phi^2 - \frac{\mu_
0^2}{\lambda'_0} \right)^2$ and $\lambda'_0=\lambda_0/6$. As is well known \cite{raja}, the Euler-Lagrange equation can be
easily obtained and is a second-order nonlinear PDE with the
following solutions: Two non-topological static solutions
$\phi_{\rm{vac.}}=:v_0=\dd \pm \mu_ 0/\sqrt{\lambda'_0}$, and two
topological static ones $\phi_{\rm{kink}}(z)=\pm \mu_ 0/\sqrt
{\lambda'_ 0}\tanh [\mu_0(z-z_0)/\sqrt 2]$ which are called kink and
antikink, respectively. The presence of an arbitrary $z_0$ is a manifestations of the translational invariance of the system, and this will lead to a zero mode. However, as mentioned before, fixing $z_0$ will break that symmetry. In
order to find the quantum corrections to this mass, we have to
make a functional Taylor expansion of the potential about the
static solutions which yields the stability equation,
\begin{equation}\label{3}
\left[ - \nabla ^2  + \frac{{d^2 U}}{{d\phi ^2 }}\bigg |_{\phi
_{\rm{static}} (z)} \right]\eta (x) = \omega^2 \eta (x),
\end{equation}
where we have defined $\phi= \phi_{\rm{static}}+\eta$ and
$\omega^{2} = k ^2 + 2\mu_0^2$. The results in the trivial sector
are the following continuum states $ \eta(x) = \exp (ikx)$. In the kink sector we
have the following two localized states and continuum states for the transparent potential in Eq.~(\ref{3}) \cite{Jackiw2}:
\begin{eqnarray}
 \eta _0 (z') &=& \sqrt {\frac{{3m_0}}{8}} \frac{1}{{\cosh^2
z'}},\nonumber\\
   \eta _B (z') &=&\sqrt {\frac{{3m_0}}{4}} \frac{{\sinh z'}}{\cosh^2
  z'},
  \nonumber \\
 \eta _q (z') & = &\frac{{e^{iqz'} }}{{N_q }}\left[ - 3\tanh^2z' + 1 + q^2  +
 3iq\tanh z'\right],
\end{eqnarray}
where $m_0=\mu_0/\sqrt{2}$, $ \omega _0 ^2  =  0$ is for our zero mode,  $ \omega _B ^2  =  \frac{3}{4}m_0^2$ is for our only bound state, and $\omega _q ^2  =
 m_0^2(\frac{q^2}{4}+1)$ are for the continuum states. Here  $N_q ^2= 16\frac{{\omega _q ^2 }}{{m_0^4 }}(\omega _q ^2  - \omega _B ^2 )$
and $z'=m_0z/2$.

\begin{figure}[th]\begin{center} \includegraphics[width=11cm]{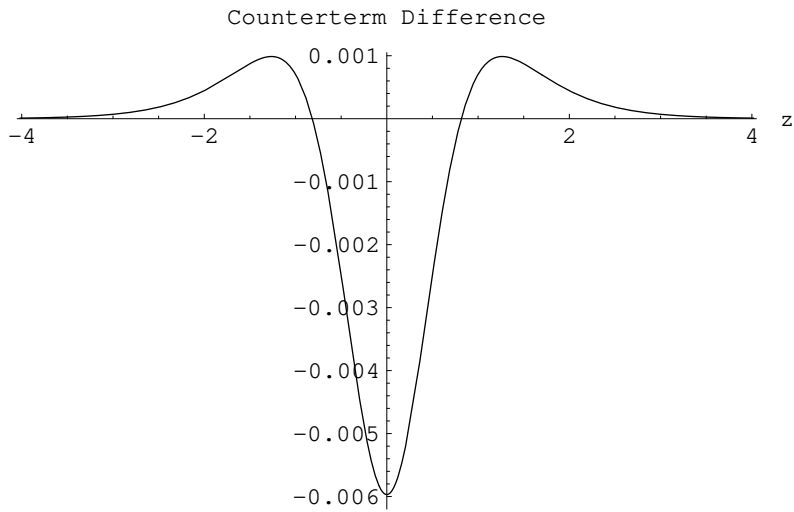}
\caption{\label{fig2}\small The difference between the counterterms for the $\phi ^4$ kink problem: $ \delta m_{\rm{kink}}^2(z)-\delta m_{\rm{free}}^2$, with mass $m=1$ and $\lambda=0.1(\mbox{m}^2)$ Note that this difference is very small and localized at $z=0$, the central position of the kink.}
\end{center}\end{figure}

Now we calculate the mass counterterm in the kink sector by
expanding the Lagrangian, which includes the mass counterterm,
around the kink background. However, we can setup a more general
problem by the following expansion
\[
\phi (z,t) \rightarrow y(z) + \eta (z,t) = \frac{m_0}{{\sqrt {2\lambda'_0}
}}\tanh[\frac{m_0}{{ 2 }}z] + \eta (z,t),
\]
where $y(z)$ can be any of the static solutions, for example the
kink solution as indicated above. Then the Lagrangian which
includes the mass counterterm becomes,
\begin{eqnarray}\nonumber\label{39}
\hspace{-.4cm} {\cal L }&=&\frac{{\rm{1}}}{{\rm{2}}}(\partial _\mu  \phi )^2  + \frac{1}{2}(m^2 - \delta m^{2})\phi ^2  - \frac{\lambda' }{4}\phi ^4  - \frac{{(m^2  - \delta m^{2})^2 }}{{4\lambda' }} \\\nonumber
 &&\hspace{-.4cm}=\frac{1}{2}(\partial _\mu  \eta )^2   + (\frac{1}{2}m^2  - \frac{3}{2}\lambda' y^2 )\eta ^2
  - \lambda' y\eta ^3  - \frac{1}{4}\lambda' \eta ^4+\delta m^{2}y\eta  \\\nonumber
  &&-\frac{1}{2}\delta m^{2}\eta ^2
   -\frac{1}{2}(\partial _\mu  y
  )^2+\frac{1}{2}(m^2-\delta m^{2})y^2- \frac{1}{4}\lambda' y ^4 \\
  &&- \frac{{(m^2  - \delta m^{2})^2 }}{{4\lambda' }}+ (m^2 y - \lambda' y^3  +
  \partial _\mu  y\partial ^\mu  )\eta.
 \end{eqnarray}
Note that the last term in the above equation which is
proportional to $\eta$ vanishes exactly after an integration by
parts and using the equation of motion. Therefore the condition of
setting the tadpole equal to zero simply becomes
\begin{eqnarray}\label{f.rules}
\raisebox{-7mm}{\includegraphics{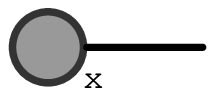}}\hspace{2mm}=
\hspace{2mm}\raisebox{-3mm}{\includegraphics{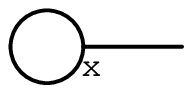}}\hspace{2mm}
+\hspace{2mm}\raisebox{-3.4mm}{\includegraphics{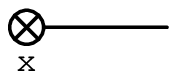}}\hspace{2mm}+\dots
=0.
\end{eqnarray}
Accordingly, up to first order in $\lambda$ we obtain,
\[
i\delta m^{2}(z,t)y(z) = \frac{1}{2}[ -i\lambda y(z)]G(z,t;z,t),
\]
where $G(z,t;z',t')$ is the propagator for the particular problem
under investigation. We finally obtain the following general
result, which is also obtained in the previous section using
analogous general arguments but a slightly different method,
\begin{equation}\label{43}
\delta m^{2}(z,t) =  - \frac{\lambda}{2} G(z,t;z,t).
\end{equation}
Note that the counterterms in general naturally turn
out to be position dependent. Since $G(z,t;z',t')$ is uniquely
determined by the nature of the problem, so is $\delta m^{2}(z,t)$
via Eq.~(\ref{43}). We do not expect any time dependence for the counterterms when the system is time translation invariant. The Green's function for this problem in the
presence of a kink is,
\begin{eqnarray}\label{28}
G(z,t;z',t') = i\int {\frac{{d\omega }}{{2\pi }}} e^{i\omega (t -
t')} \left(\sum\limits_{n \ne 0} {\frac{{\eta ^* _n (z)\eta _n
(z')}}{{\omega^2  _n  - \omega^2 }}}  + \int {dk} \frac{{\eta ^*
_k (z)\eta _k (z')}}{{\omega^2  _k - \omega^2  }}\right),
\end{eqnarray}
where the sum indicates the contributions of the bound states and
the integral the continuum states. Note that the zero mode is
neglected since it is only a manifestation of the translational
invariance of the system and is to be treated as a collective
coordinate \cite{raja,shif}. The above equation, when the two
space-time points are set to be equal and the $\omega$ integration
is performed, becomes
\begin{equation}\label{30.5} G(z,t;z,t)
=-\frac{{\eta _B ^2 (z)}}{{2\omega_B }}  - \int {\frac{{dk}}{{2\pi
}}} \frac{{\left| {\eta_k (z)} \right|^2 }}{{2\omega_k }}.
\end{equation}
Calculating this integral is very cumbersome, but we can use an
interesting relationship which is the local version of the
completeness relation \cite{gousheh2,gold,reb2002}:
\begin{equation}\label{31}
{\left| {\eta_k(z)} \right|^2  = 1 - \frac{m}{{\omega_k ^2  -
\omega_B ^2 }}\eta _B ^2 (z) - \frac{{2m}}{{\omega_k ^2 }}\eta _0
^2 (z)}.
\end{equation}
Using the above equation, Green's function is easily computable by
performing simple integrals. Putting Eq.~(\ref {31}) into Eq.~(\ref
{30.5}) and using Eq. (\ref{43}), the mass counterterm in the kink
background becomes,
\begin{equation}\label{32}
\delta m^{2}_{\rm{kink}}(z)  = \frac{\lambda }{{6\sqrt 3 m}}\eta _B ^2
(z) - \frac{{\lambda }}{{2\pi m}}\eta _0 ^2 (z) + \frac{{\lambda
}}{{8\pi }}\int_{- \infty }^\infty  {dk\frac{1}{{\sqrt {k^2  + m^2
} }}},
\end{equation}
which as expected earlier is different from mass counterterm in
the trivial sector, i.e. the last term in Eq.~(\ref{32}). In fact
it has extra finite $z$-dependent terms due to the presence of the
localized states, and obviously this difference tends to zero as
$z\to\pm\infty$. An alternative reasoning is that the kink
solution also tends to either of the trivial vacuum states as
$z\to\pm\infty$. Therefore the  difference between the counterterms becomes
\begin{equation}\label{32}
\delta m^{2}_{\rm{kink}}(z) -\delta m^{2}_{\rm{free}} = \frac{\lambda }{{6\sqrt 3 m}}\eta _B ^2
(z) - \frac{{\lambda }}{{2\pi m}}\eta _0 ^2 (z).
\end{equation}
Note that the zero mode appears in the above expression due to the completeness relationship, Eq.~(\ref{32}). However it dose not appear in the Green's function or the field quantization. Figure \ref{fig2} illustrates this difference for  $m=1$ and $\lambda=0.1(\mbox{m}^2)$, where m denotes the mass of the scalar particle. We have computed the Casimir energy for this system using this procedure \cite{mohammadi}

\section{Conclusions}
In this paper we have studied the effects of nontrivial boundary
conditions or backgrounds on the renormalization program for a
given system in two dimensions. In general these two non-perturbative conditions on the system are
non-perturbative effects. In the case of nontrivial backgrounds
the effect is certainly non-perturbative when the boundary
conditions at infinities are altered, e.g. in the presence of
solitons. We have insisted that the renormalization program could
completely take into account the boundary
conditions or any possible nontrivial backgrounds which break the
translational invariance of the system. We have shown that the
problem can be self-contained and the above program is accomplishable.
To be more specific, we believe, in principle there should be no need to import
counterterms from the free theory, or even supplementing them with
the \emph {ad hoc} attachment of extra surface terms, to remedy
the divergences inherent in this theory. In general this breaking
of the translational invariance, is reflected in the nontrivial
position dependence of all the n-point functions. As we have
shown, this could have profound consequences. For example in the case of
renormalized perturbation theory, the counterterms and hence the
radiative corrections to parameters of the theory, i.e. $m$ and
$\lambda$, automatically turn out to be position dependent in our approach. In
particular we have calculated the mass
counterterms for both the nontrivial boundary condition and nontrivial background cases, and computed and plotted their differences with the free case in figures \ref{fig1} and \ref{fig2}. We have used this renormalization program to compute the Casimir energies in some cases \cite{Gousheh1,Gousheh2,mohammadi}.

\vskip20pt\noindent {\large {\bf
Acknowledgements}}\vskip5pt\noindent

This research was supported by the research office of the Shahid Beheshti University.

\end{document}